# Acoustic Vortex in Waveguide with Chiral Gradient Sawtooth Metasurface

Zeliang Song [1], Shuhuan Xie [2], Yong Li [2], Hua Ding [2], Feiyan Cai [3], Yugui Peng [1], Xuefeng Zhu [1] and Degang Zhao [1,*]

[1] Department of Physics, Huazhong University of Science and Technology, Wuhan 430074, China.

[2] Institute of Acoustics, School of Physics Science and Engineering, Tongji University, Shanghai 200092, China.

[3] Paul C. Lauterbur Research Center for Biomedical Imaging, Shenzhen Institutes of Advanced Technology, Chinese Academy of Sciences, Shenzhen 518055, China.

* Corresponding author. E-mail address: dgzhao@hust.edu.cn (D. Zhao).

**Abstract**

The acoustic vortex states with spiral phase dislocation that can carry orbital angular moment (OAM) have aroused many research interests in recent years. The mainstream methods of generating acoustic vortex are based on Huygens-Fresnel principle to modulate the wavefront to create spatial spiral phase dislocation. In this work, we propose an entirely new scenario to generate acoustic vortex in a waveguide with chiral gradient sawtooth metasurface. The physical mechanism of our method is to lift the degenerate dipole eigenmodes through the scattering effect of the chiral surface structure, and then the superposition of them will generate both $+1$ and $-1$ order vortices in place. Compared to the existing methods of acoustic vortex production, our design has many merits, such as easy to manufacture and control, the working frequency is broadband, sign of vortex order can be readily flipped. Both the full-wave simulations and experimental measurements validate the existence of the acoustic vortices. The torque effect of the acoustic vortices is also successfully performed by rotating a foam disk as a practical application. Our work opens up a new route for generating acoustic vortex and could have potential significances in microfluidics, acoustic tweezers and ultrasonic communication, etc.

**Introduction**

Sound waves with orbital angular momentum (OAM), called acoustic vortex, have opened up a brand-new approach of acoustic wave controlling and aroused lots of research interests. The pressure field of the acoustic vortex behaves helicoidal phase dependence in wavefront, which can be described by a phase term $e^{il\varphi}$. $l$ is the topological charge or the order of the vortex, representing the number of twists in a wavelength along propagating direction or the number of $2\pi$ phase cycles around a closed path in wavefront. The concept of this kind of phase dislocation was initially proposed by Nye and Berry as early as 1974 [1]. The vortex wave contains phase singularity (where the phase becomes indeterminate) at the center axis, accompanied by a null pressure magnitude[1-3]. This peculiar characteristic can be used in non-contact particle manipulation such as particle trapping [4-16]. On the other hand, the acoustic vortex wavefront can exert torque on the sound absorbing objects placed on the propagation path. This torque effect can be used for non-contact rotation manipulation [17-28]. In addition, the distinct topological charges of OAM provide orthogonal channels that serve as an extra independent degree of freedom in transmitting information. Therefore, the acoustic vortex has the potential capability of acoustic communication through encoding information onto the vortex waves [29-33].

Over the past two decades, some significant achievements have emerged for generating acoustic vortices. These methods can be generally classified into two types: active and passive. The active method refers to straightforwardly producing the helicoidal phase by using phase-controllable sound source array, which was firstly proposed in Hefner and Marston's pioneer work using 4 panel transducers with gradient varied phases [2,3]. Since then, high-quality acoustic vortex beam can be generated by using state-of-the-art transducer arrays applying more and more transducers and exactly presetting each transducer's phase and amplitude which are precisely controlled by digital processors [4-7,17-20,33-39]. Technically speaking, the active method can efficiently generate acoustic vortices for any topological charge and any allowed frequencies. Nonetheless, the cost of large number transducers is very high and the control system is inevitably complicated. The passive method relies on the well-designed artificial microstructures to modulate the sound field. When a sound plane wave passes through the microstructures, its uniform phase on the wavefront can be dislocated to a spatial helical distribution for some given frequencies. This passive technology was firstly realized using structures with helical dislocation surfaces [40,41]. Subsequently, a variety of inventive and ingenious designs have been proposed and developed, such as Helmholtz resonator-based metasurfaces [21,22,42-44], spiral diffraction gratings [10,45-49], acoustic holographic lens [50,51], non-coaxial waveguide [52], and so on. This kind of method is comparatively inexpensive and the structure is usually easy to fabricate. However, the topology architectures of microstructures heavily depend on the working frequency and generally one structure has only one working frequency instead of a frequency range.

In essence, the underlying physics of these two commonly accepted methods is the Huygens-Fresnel principle. While in this work, we put forward an entirely new approach on generating acoustic vortices using a single sound source in a cylindrical waveguide, which is designed with sawtooth-like grooves that are evenly-increased distributed along the circumferential direction on the surface. These azimuthal gradient grooves can remove the degeneracy of waveguide's eigenmodes, which enables them to have different wave number $k_z$ in propagating direction under the same excited frequency. Owing to the difference of growth of phase term $e^{ik_z z}$, the superposition of the two degeneracy-lifted modes can form vortex fields wherever their phase difference equals to an odd number times $\pm\pi/2$. Consequently, in a wide frequency region, two vortex wave fields with topological charge $l = \pm 1$ can be generated, and the topological charge can be readily flipped by adjusting the position of sound source on the incident plane, which are demonstrated by full-wave simulation obtained using finite-element solver COMSOL Multiphysics and verified by experimental measurements. Furthermore, to visually exhibit the acoustic vortex field, we show the torque effect on an absorbing foam disk due to the transferring of OAM from vortex beam to matter. Our work offers a new mechanism that relies on superposition of waveguide's eigenmodes to generate acoustic vortex field and provide a new way to design economic, broadband, multifunctional acoustic vortex devices.

**Results**

To start with, we consider a typical solid infinitely long circular waveguide of radius $r_0 = 40\,\text{mm}$ filled with air, as shown in Fig. 1a. By solving the Helmholtz equation in cylindrical coordinates $(r, \varphi, z)$, the sound pressure eigensolutions can be expressed as [53-55]

$$p_{mn} = J_m(k_{mn}r)(A_m \cos m\varphi + B_m \sin m\varphi)e^{ik_z z}, \qquad (1)$$

where $J_m(k_{mn}r)$ is the first kind Bessel function of $m$th order $(m \geq 0)$. $k_{mn}$ is the in-plane wave vector component and it can be determined by the Dirichlet boundary condition $[\partial J_m(k_{mn}r)/\partial r]_{r=r_0} = 0$, where $n$ denotes the $n$th root. $k_z$ refers to the propagation wave vector component along the $z$ axis. It satisfies $k_{mn}^2 + k_z^2 = \omega^2/c^2$, where $\omega = 2\pi f$ is the angular frequency and $c = 343$m/s is the sound velocity in the air. Then each eigenmode can be labeled by two integers $(m,n)$.

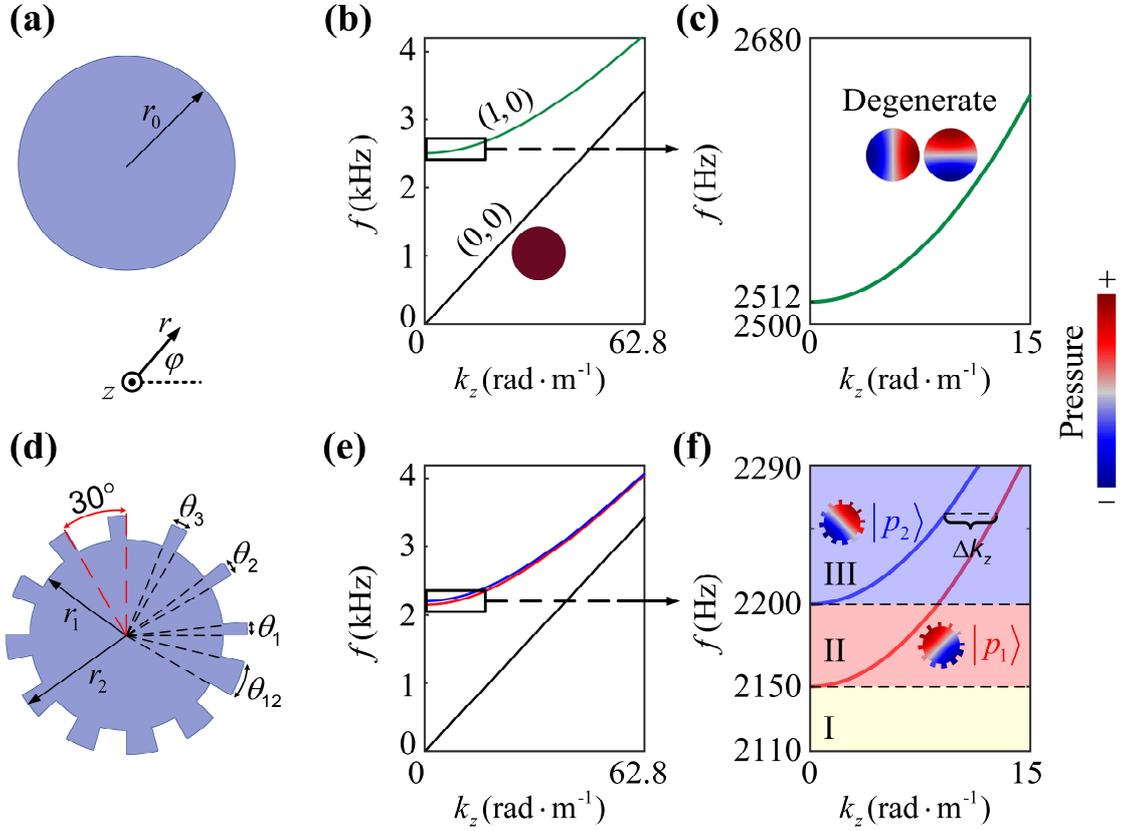

FIG. 1. Eigenmodes analysis and degeneracy splitting. a) Sectional view of circular waveguide. Radius $r_0 = 40$mm. The waveguide is infinite long in $z$ direction. b) Dispersion relations of two lowest-order eigenmodes of circular waveguide. The insert is the pressure eigenfield distribution of $(0,0)$ mode. c) Magnified view of dispersion relation of the $(1,0)$ mode in (b). The inserts are the pressure eigenfield distributions of degenerate $(1,0)$ mode. d) Sectional view of the waveguide with chiral gradient sawtooth metasurface. All grooves are homocentric fans, with inner radius $r_1 = 40\text{mm}$ and outer radius $r_2 = 50\text{mm}$. The structure consists of 12 pairs of grooves in total. The central angle of each convex-concave pair is fixed to $30°$, while that of convex part follows the function $\theta_i = \theta_1 + (i-1)\delta$, $i = 1, 2, \cdots, 12$ with $\theta_1 = 6°$ and $\delta = 1°$. e) Dispersion relations of the three lowest-order eigenmodes of the waveguide with cross section shown in (d). f) Magnified view of the degeneracy-lifted modes in (e). The inserts are the pressure eigenfield distributions of $|p_1\rangle$ and $|p_2\rangle$ modes at the cut-off frequencies. At a certain frequency above the higher cut-off frequency $2200\text{Hz}$, the horizontal interval of the two dispersion curves denotes the difference of $k_z : \Delta k_z = k_{1z} - k_{2z}$.

The dispersion relation (the relation between eigenfrequency and $k_z$) of this cylindrical waveguide is depicted in Fig. 1b. The lowest straight line represents the $(0,0)$ mode satisfying $\omega = ck_z$ and it is actually the plane wave component $A_0|p_0\rangle = A_0 e^{ik_z z}$. Its pressure field is uniformly distributed in the transverse plane, which is known as the monopole mode (see the inset of Fig. 1b). The second curve represents $(1,0)$ mode, whose cut-off frequency is $\frac{c}{2\pi}k_{10} = 1.841\frac{c}{2\pi r_0} = 2512\text{Hz}$, which can be exactly determined by solving boundary condition [53-55]. As shown in Fig. 1c, the field distribution clearly reveals that they are two orthogonal dipole modes [38,52,56]. Henceforth, we mainly focus on $(1,0)$ modes and the dispersion curves of higher order modes are not given here. However, in the circular waveguide, the vortex mode can't be generated by common sound source such as plane wave or single point source directly, because the two definitely degenerate dipole modes have no phase difference and their superposition can not bring about gradient phase term $e^{il\varphi}$. That's why the previous works must utilize well-designed sound source array or artificial microstructures to produce phase gradient intentionally [38,39,42,57].

It is natural to ask a question: whether these two degenerate modes can be lifted? Geometrically the degeneracy feature of the dipole eigenmodes originates from the spatial inversion symmetry in the azimuthal direction. If an artificial structure is elaborately designed on the surface to break this symmetry, it is anticipated that the degeneracy can be lifted. On these grounds, we carve 12 grooves with gradient size on the surface of waveguides to break the circumferential symmetry, as shown in Fig. 1d. The width of grooves is $r_2 - r_1 = 10\text{mm}$, which is about one order smaller than the working wavelength. Then this groove structure can act as a metasurface. The central angle of each convex-concave pair is fixed as $30°$, while the central angles of convex part linearly increase anticlockwise from $\theta_1 = 6°$ to $\theta_{12} = 17°$ with step $1°$. The numerically calculated dispersion relation and eigenfield distribution of pressure are presented in Figs. 1e and 1f. It is obvious that the original two degenerate dipole modes split into two distinguishable modes. The new cut-off frequencies are $f_1 = 2150\text{Hz}$ and $f_2 = 2200\text{Hz}$, respectively. These two degeneracy-lifted eigenmodes can be approximatively expressed as

$$|p_1\rangle \approx J_1(k_{1r}r)e^{ik_{1z}z}\sin(\varphi - \varphi_0)$$
$$|p_2\rangle \approx J_1(k_{2r}r)e^{ik_{2z}z}\cos(\varphi - \varphi_0)$$
(2)

where $\varphi_0 = 44°$ determines the polarization direction of the dipole modes. It should be pointed out that these degeneracy-lifted modes have definite $\varphi_0$, while for the degenerate modes in circular waveguide $\varphi_0$ can be any value [55] and we set $\varphi_0 = 0$ in Fig. 1c for simplicity. The patterns of eigenfield distributions in the transverse plane for $|p_1\rangle$ and $|p_2\rangle$ modes, as shown in the insets of Fig. 1f, are nearly unchanged compared with the degenerate modes in circular waveguide, only having very little deviation near the boundary. Then the approximation in Eq. (2) is absolutely valid and the detailed analysis about the validity of approximation can be found in Supplemental Materials S1. In principle, the chiral gradient sawtooth metasurface effectively introduces small perturbation that successfully lifts the two degenerate modes while preserving their wave function relation. The similar phenomenon is frequently encountered in quantum system, which can be analytically solved by well-established perturbation theory. It should be noted that the higher degenerate modes can also be lifted which has been discussed in Supplemental Materials S2.

The lift of the degeneracy divides the frequency domain into three zones, which are revealed by different color-filled regions in Fig. 1f. A sound source with frequency in zone I will only excite the monopole mode $|p_0\rangle$. In zone II, the dipole mode $|p_1\rangle$ can be excited and it coexists with the monopole mode. In zone III, $|p_1\rangle$, $|p_2\rangle$ and monopole mode can be excited simultaneously because the excited frequency exceeds $f_2$ (here the working frequency is limited below the cut-off frequency of the eigenmode higher than $|p_2\rangle$ to avoid exciting higher order eigenmodes). Thereupon the pressure field in any place of waveguide can be expressed as the linear superposition of eigenmodes $|p\rangle = A_0|p_0\rangle + A_1|p_1\rangle + A_2|p_2\rangle$. Through ingenious design, the acoustic vortex can be generated when two essential conditions are satisfied:

$$(a) \text{ Amplitude condition: } |A_1 J_1(k_{1r} r)| = |A_2 J_1(k_{2r} r)| = A.$$
$$(b) \text{ Phase difference condition: } \Delta k_z z = (k_{1z} - k_{2z})z = (2N-1)\pi/2, N \in \mathbb{N}. \tag{3}$$

For example, assuming $-A_1 J_1(k_{1r} r) = A_2 J_1(k_{2r} r) = A$ satisfying the condition (*a*), $\Delta k_z z = \pi/2$ satisfying the condition (*b*), the superposition of states $|p_1\rangle$ and $|p_2\rangle$ is

$$\begin{aligned} A_1|p_1\rangle + A_2|p_2\rangle &= A_1 J_1(k_{1r} r)\sin(\varphi - \varphi_0)e^{ik_{1z}z} + A_2 J_1(k_{2r} r)\cos(\varphi - \varphi_0)e^{ik_{2z}z} \\ &= A e^{ik_{2z}z} \left[ e^{i\Delta k_z z} \sin(\varphi - \varphi_0) + \cos(\varphi - \varphi_0) \right] \\ &= A e^{ik_{2z}z} \left[ i\sin(\varphi - \varphi_0) + \cos(\varphi - \varphi_0) \right] \\ &= A e^{i(k_{2z}z + \varphi - \varphi_0)} \end{aligned} \tag{4}$$

The gradient phase term $e^{i\varphi}$ in Eq. (4) indicates the helical phase distribution and a vortex mode with topological charge $l = +1$ has been produced successfully. It should be noted that although the monopole mode $|p_0\rangle$ coexists, fortunately its field is uniformly distributed in the wavefront and it does not provide any in-plane phase gradient but only acts as a background.

In principle, the amplitude condition Eq. 3($a$) depends on the modal amplitudes $A_1$ and $A_2$, or in other words the relative weights of modes $|p_1\rangle$ and $|p_2\rangle$. A simple plane wave source should be located off-center and the modal amplitudes can be readily adjusted only by tuning the azimuthal angle $\varphi_s$ of the source. Detailed mathematical derivation are presented in Supplemental Materials S3. While the phase difference condition Eq. 3($b$) relies on $\Delta k_z$ (determined by dispersion relation) and the distance $z$ from the incident plane.

To demonstrate our strategy for creating the acoustic vortex, we fabricated the waveguide samples according the parameters in Fig. 1d via stereolithographic 3D printing technology, as shown in Fig. 2a. The raw material of the sample is photosensitive resin and its wall thickness is $d = 3\text{mm}$ to guarantee the hard boundary condition. As is shown in Fig. 2b, a loudspeaker with diameter of about $16\text{mm}$ is off-center placed in the input port, with the offset distance is about $r_s = 25\text{mm}$. The loudspeaker is embedded in a sound-h). The parameter settings for three cases are: (c) and (f), $\varphi_s = -7°$, $z = 0.61\mathrm{m}$; (d) and (g), $\varphi_s = 95°$, $z = 0.61\mathrm{m}$; (e) and (h), $\varphi_s = -7°$, $z = 1.83\mathrm{m}$. The working frequency is fixed to be $2330\mathrm{Hz}$.

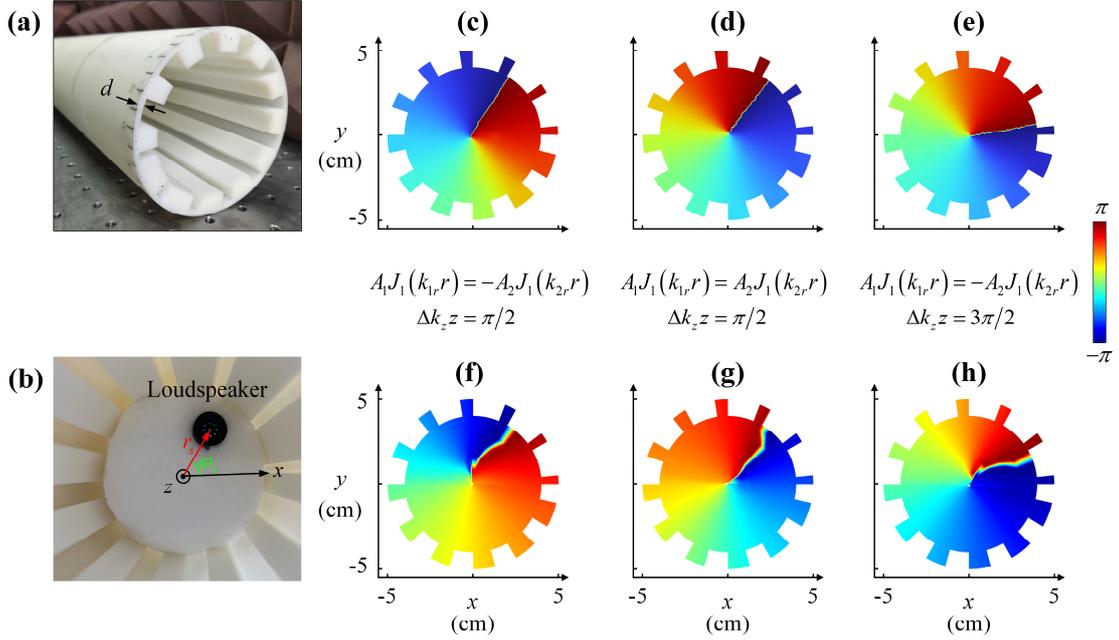

FIG. 2. Phase distribution of the acoustic vortex. a) Photo of the fabricated waveguide sample. b) Experimental setup of loudspeaker. c-h) Phase distribution of acoustic vortex for three cases obtained by numerical simulation (c-e) and experimental measurement (f-

absorbing foam layer, which is used to stick loudspeaker and the azimuthal angle of the loudspeaker can be easily adjusted by manually rotating the foam plate. The output port of the waveguide is full filled by another sound-absorbing foam to playing the role of a perfectly matched layer (PML). A 1/8-inch microphone is inserted into the waveguide to scan the sound pressure. Here we choose the working frequency as $2330\mathrm{Hz}$, at which $\Delta k_z = 2.58\ \mathrm{rad/m}$ obtained by the dispersion relations shown in Fig. 1e. Firstly, we set the azimuthal angle of sound source as $\varphi_s = -7°$ which satisfies $-A_1 J_1(k_{1r} r) = A_2 J_1(k_{2r} r)$, and meanwhile the examined plane is chosen at $z = 0.61\mathrm{m}$ which satisfies $\Delta k_z z = \pi/2$. This setting will produce a +1 order pure acoustic vortex, as previously discussed in Eq. (4). The phase distributions of simulation and experiment are exhibited in Figs. 2c and 2f, respectively. They match very well and clearly show that the phase anticlockwise increase along the azimuthal direction and jumps $2\pi$ over one annular loop, revealing a vortex mode with topological charge $+1$. Next we adjust $\varphi_s$ to be $95°$ and the amplitude condition becomes $A_1 J_1(k_{1r} r) = A_2 J_1(k_{2r} r)$, and keep the position of examined plane unchanged. Using the derivation similar to Eq. (4), a phase term $e^{-i\varphi}$ will appear, implying a $-1$ order acoustic vortex has been produced. The well-matched simulated (Fig. 2d) and experimental (Fig. 2g) phase distributions perfectly demonstrate the existence of $-1$ order vortex: the phase clockwise increases $2\pi$ over one annular loop. At the same frequency, the acoustic vortex can also be obtained at the different positions on the z-axis. For instance, we fix $\varphi_s = -7°$, i.e. $-A_1 J_1(k_{1r} r) = A_2 J_1(k_{2r} r)$, and the examined plane is changed at $z = 1.83\mathrm{m}$, which satisfies $\Delta k_z z = 3\pi/2$. Calculating $A_1|p_1\rangle + A_2|p_2\rangle$ will obtain $e^{-i\varphi}$ phase term, thus the vortex with topological charge $-1$ will appear. The phase distribution in this case is similar to those shown in Figs. 2d and 2g, and the numerical simulation (Fig. 2e) agrees well with the experimental measurement (Fig. 2h). Certainly, the vortex mode can also exist at the position that satisfies $\Delta k_z z = 5\pi/2, 7\pi/2, 9\pi/2, \cdots$. However, in experiment it is not easy to obtain high-quality vortex at the distance too far from the source, which is due to the inevitable dissipation of the propagating acoustic wave.

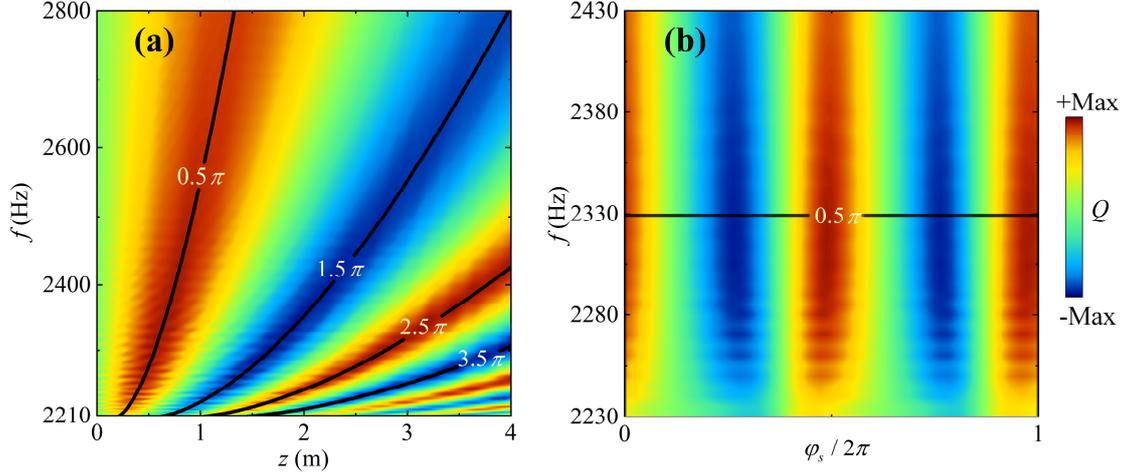

FIG. 3. Quality of the vortex. a) The $Q$ factor of the vortex states versus frequency and $z$ coordinate of examined plane, with the sound source is fixed at $\varphi_s = -7°$. The red and blue colors denote the $+1$ and $-1$ order vortex, respectively. b) The $Q$ factor of the vortex states versus frequency and the azimuthal angle $\varphi_s$ of sound source, with the examined plane is fixed at $z = 0.61 \text{m}$. The solid curves are plotted according to the phase difference condition Eq. 3(*b*), and the numbers in the curves denote the quantity of $\Delta k_z z$.

To describe the quality of the vortex, a $Q$ factor $Q = \dfrac{\text{Im}\langle p|\partial_\varphi|p\rangle}{\langle p|p\rangle}$ can be defined [58,59], which essentially indicates the expectation value of the OAM in the state $|p\rangle$. In Fig. 3a, we numerically calculate $Q$ factor with respect to frequency and $z$ coordinate of the examined plane, with sound source is fixed at $\varphi_s = -7°$. The vortices having the best quality are on the curves calculated by $\Delta k_z z = (2N-1)\pi/2, N \in \mathbb{N}$, i.e., the phase difference condition Eq. 3(*b*), while the quality of vortex declines away from the curves. Obviously, in our design only one waveguide sample can achieve high quality vortices in a very wide frequency region instead of only single working frequency. According to Fig.

3a, at different frequencies, high quality vortices can be obtained at different $z$ coordinate for both +1 (red region) and −1 (blue region) orders. Fig. 3b presents the $Q$ factor calculated in the transverse plane at $z = 0.61\text{m}$ versus the frequency and the azimuthal angle $\varphi_s$ of sound source. Similar to Fig. 3a, the vortex states having best quality locate on the line $\Delta k_z z = \pi/2$. In a $2\pi$ period of $\varphi_s$, the $Q$ factor exhibits a cyclical variation, and the sign of the topological charge of vortex flips every $\pi/2$ change of $\varphi_s$. This periodicity originates from the periodic functional relationship of the relative weights of $|p_1\rangle$ and $|p_2\rangle$ with respect to $\varphi_s$, which have been demonstrated in Supplemental Materials S3.

One fantastic effect of acoustic vortex is that the spiral field can transfer OAM to a sound absorbing object, which is known as the vortex-induced torque effect [18,21-24]. Here we experimentally demonstrate this torque effect by non-contact rotating a light object. As schematically shown in Fig. 4a, a loudspeaker is off-centered placed at the bottom of a waveguide to act as the sound source, with working frequency $2330\text{Hz}$. A light sound absorbing foam disk is hung $0.61\text{m}$ above the loudspeaker. A camera is fixed above the foam disk to take videos. Firstly, the loudspeaker is placed at $\varphi_s = -7°$, and as soon as the sound is launched from loudspeaker, the foam disk starts rotating clockwise, as is shown in the top row of Fig. 4b due to the +1 order vortex. The multimedia file (video1.mp4) in Supplemental Materials visually displays the rotational motion of the disk. Here the selective hang thread has very small shear modulus, thus it produces very small torque when it is twisted. Therefore, the disk can continuously rotate for a very long time. As is discussed in Fig. 3b, the topological charge of vortex can be easily flipped by tuning the azimuthal angle of sound source. Then we place the loudspeaker at $\varphi_s = 95°$, unsurprisingly the disk rotates anticlockwise due to the −1 order vortex has been excited, which is clearly shown in the bottom row of Fig. 4b and the multimedia file (video2.mp4) in Supplemental Materials. In addition, we fix the loudspeaker on a turntable and rotate the turntable one full round. It intuitively shows that the rotational direction of the disk flips four times, which is displayed in video3.mp4 in Supplemental Materials.

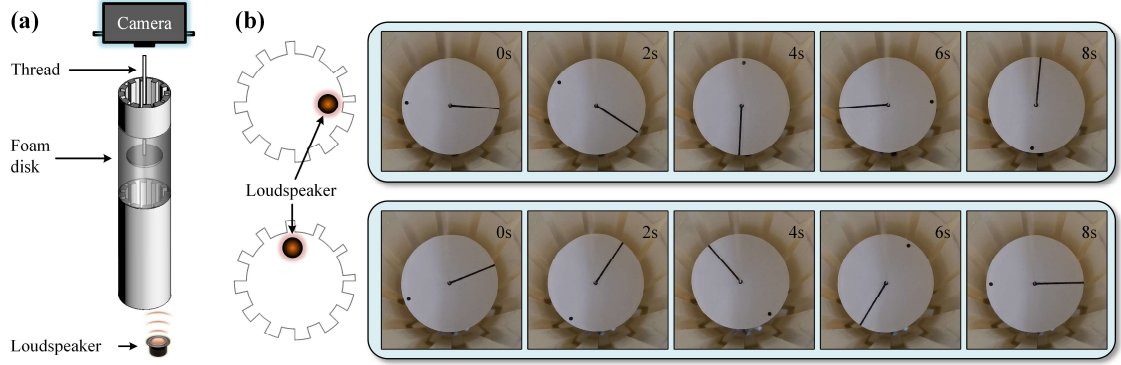

FIG. 4. The torque effect of acoustic vortex exerted on an absorbing foam disk. a) Schematic diagram of experimental setup. b) Snapshots of the rotational motion of the foam disk at different time points. The foam disk is placed at $z = 0.61\text{m}$ away from the incident plane. A black line segment is marked for the observation of the disk rotation. The loudspeaker is placed at $\varphi_s = -7°$ (top) and $\varphi_s = 95°$ (bottom), and consequently the foam disk rotates clockwise (top) and anticlockwise (bottom).

**Discussion**

In this work, we have proposed a brand-new novel approach on generating acoustic vortex in a cylindrical waveguide with chiral gradient sawtooth metasurface. The chiral surface structure is used to split the degenerate dipole eigenmodes. The superposition of these two orthogonal modes will produce ±1 order acoustic vortices, as long as they satisfy two essential conditions: amplitude condition and phase difference condition. The phase distribution measurement and torque effect experiment exactly verify the validity of our theory. The key point of our method is the lift of the degenerate dipole modes, which is achieved by the scattering effect of the acoustic wave on the metasurface structure. Certainly, the configuration of metasurface structure is not unique. Other kinds of surface design may realize similar effect. However, the architecture of the metasurface structure should be designed very carefully: it should not break the functional relations of dipole modes very much, i.e., Eq. (2) should be satisfied, and simultaneously, the two dipole modes should have enough difference of the propagating wave number $\Delta k_z$. In this work, if the size of the sawtooth is too big or the gradient of size is too great, although the wave number difference $\Delta k_z$ may become wider, the dipole property of eigenmodes may be badly distorted by the strong wave scattering on the surface and high-quality vortex cannot be obtained. Although it is well known that the acoustic wave is a longitudinal wave, the acoustic dipole modes in waveguide have some characteristics of transverse waves: they have wave vector components $k_r$ which is perpendicular to the propagation direction [38,53]. From this point of view, the two orthogonal dipole modes $|p_1\rangle$ and $|p_2\rangle$ can be analogized as TE and TM modes in optics. Then the physical mechanism of generating acoustic vortex in our work is similar to the generating of circularly polarized light: passing a linearly polarized light through a quarter-wave plate at an angle of 45° to the optic axis of the plate. In contrast to conventional methods for generating sound vortices that rely on complex electronic control or intricate craftmanship, our design offers many advantages. Our structure is simple and easy to fabricate. We can use only one waveguide and only one loudspeaker to generate vortices in broadband frequency region. According to the phase difference condition, the vortices of different frequencies can be obtained at the transverse plane at different distances from the sound source. And in terms of the amplitude condition, the sign of topological charge and the quality of vortex can be flexibly tuned through adjusting the angular position of the sound source. Our study provides a new thought of generating sound vortex and could be potentially significant in many practical applications, including but not limited to micro-mixer in microfluidic system, non-contact manipulation of small objects in biology, and so on.

**Materials and Methods**

**Simulations.** Full-wave simulations are carried out by using the acoustic module of COMSOL Multiphysics, a commercial finite-element solver. The waveguide is filled with air, whose sound speed $c_0 = 343$m/s and density $\rho_0 = 1.29$kg/m$^3$. The wall of waveguide is set hard boundary condition. We set Floquet periodic boundary condition in $z$ direction, and the dispersion curves in Fig. 1 are obtained by conducting parameter sweeping of $k_z$ in eigenfrequency study. For Fig. 2, at the input plane, an off-centered small-hole (positioned $r_s, \varphi_s$) is set up as a sound source that emits sound plane wave. The phase distribution diagrams are obtained using frequency domain study in a finite length waveguide with a perfectly matched layer (PML) adding in the output plane. The data in Fig. 3 is obtained by post-processing of the simulated results, which requires doing inner product $\dfrac{\operatorname{Im}\langle p | \partial_\varphi | p \rangle}{\langle p | p \rangle}$ for all cut-planes with $\varphi_s = -7°$ (Fig. 3a) or all $\varphi_s$ at one cut-plane $z = 0.61\text{m}$ (Fig. 3b) as well as for all frequencies.

**Sample fabrication.** The sample is fabricated via stereolithographic 3D printing technology. The raw material of the sample is photosensitive resin and its wall thickness is $d = 3\text{mm}$, which is thick enough to guarantee the hard boundary condition. We fabricate 4 identical waveguide pieces with each length 0.5m. They can be easily assembled and disassembled manually. Two pieces are used in the experiment of propagation distance $z = 0.61\text{m}$, while four pieces are used in the experiment of $z = 1.83\text{m}$.

**Experimental setup and measurement.** For the phase distribution experiment, the loudspeaker is embedded in a round sound-absorbing foam plate (as shown in Fig. 2d), which is used to stick loudspeaker and the azimuthal angle of the loudspeaker can be easily adjusted by manually rotating the plate. The loudspeaker is stimulated by a waveform generator (RIGOL DG1032). The end of the waveguide is full filled by another sound-absorbing foam to play the role of PML. A 1/8-inch microphone (1/8-in. Brüel & Kjær 4138-A-015), which is tied to a long specially-made auxiliary steel rod, is horizontally moved into waveguide to scan the sound pressure. The position of measuring microphone is controlled by a Cartesian coordinate robot. During the measurement, the collected signals are sent to the channel analyzer (Brüel & Kjær 3160-A-042) to extract the phase data.

For the torque effect experiment, the waveguide sample is settled upright on a specially made middle-hollow bracket. A loudspeaker is off-centered fixed on a turntable and placed at the bottom of waveguide. A power amplifier (Aigtek ATA-304) is used between to enhance the sound intensity. A light sound absorbing foam disk (with radius 35mm and thickness 10mm) is hung 0.61m above the loudspeaker inside the waveguide by a thin soft cotton thread. Here the selective hang thread has very small shear modulus, thus it produces very small torque when it is twisted. Therefore, the disk can continuously rotate for a very long time.

**Acknowledgments**

Z. S. and D. Z. thank Mr. Xinchao Song for the helpful suggestion. This work was supported by the National Key R&D Program of China (Grant No. 2020YFA0211400) and the Opening Project of the Key Lab of Health Informatics of Chinese Academy of Sciences (Grant No. 2011DP173015-2020-1).

**Author contributions**

Z. S. carried out the theoretical analysis and numerical simulation. Z. S., S. X. and H. D. fabricated the sample and performed the measurement. D. Z. conceived the idea and supervised the study. Y. L., F. C. provided assistance in experiments. All the authors contributed to discussion, interpreting the data and the writing.

**Conflict of interest**

The authors declare that they have no conflict of interest.

**References**

[1] J. F. Nye and M. V. Berry, Dislocations in wave trains, Proc. Math. Phys. Eng. Sci. **336**, 165 (1974).
[2] B. T. Hefner and P. L. Marston, Acoustical helicoidal waves and Laguerre-Gaussian beams: Applications to scattering and to angular momentum transport, J. Acoust. Soc. Am **103**, 2971 (1998).
[3] B. T. Hefner and P. L. Marston, An acoustical helicoidal wave transducer with applications for the alignment of ultrasonic and underwater systems, J. Acoust. Soc. Am. **106**, 3313 (1999).
[4] S. T. Kang and C. K. Yeh, Potential-well model in acoustic tweezers, IEEE Trans. Ultrason. Ferroelectr. Freq. Control **57**, 1451 (2010).
[5] A. Marzo, S. A. Seah, B. W. Drinkwater, D. R. Sahoo, B. Long, and S. Subramanian, Holographic acoustic elements for manipulation of levitated objects, Nat. Commun. **6**, 8661 (2015).
[6] C. R. Courtney, C. E. Demore, H. Wu, A. Grinenko, P. D. Wilcox, S. Cochran, and B. W. Drinkwater, Independent trapping and manipulation of microparticles using dexterous acoustic tweezers, Appl. Phys. Lett. **104**, 154103 (2014).
[7] D. Baresch, J. L. Thomas, and R. Marchiano, Observation of a Single-Beam Gradient Force Acoustical Trap for Elastic Particles: Acoustical Tweezers, Phys. Rev. Lett. **116**, 024301 (2016).

[8] P. Liu, D. Ming, C. S. Tan, and B. Lin, Acoustic trapping with 3-D manipulation, Appl. Acoust. **155**, 216 (2019).

[9] M. Baudoin *et al.*, Spatially selective manipulation of cells with single-beam acoustical tweezers, Nat. Commun. **11**, 4244 (2020).

[10] N. Jiménez, V. Romero-García, L. M. García-Raffi, F. Camarena, and K. Staliunas, Sharp acoustic vortex focusing by Fresnel-spiral zone plates, Appl. Phys. Lett. **112**, 204101 (2018).

[11] S. Guo, X. Wang, X. Guo, Z. Ya, P. Wu, A. Bouakaz, and M. Wan, Decreased clot debris size and increased efficiency of acoustic vortex assisted high intensity focused ultrasound thrombolysis, J. Appl. Phys. **128**, 094901 (2020).

[12] M. Baudoin, J.-C. Gerbedoen, A. Riaud, O. B. Matar, N. Smagin, and J.-L. Thomas, Folding a focalized acoustical vortex on a flat holographic transducer: Miniaturized selective acoustical tweezers, Sci. Adv. **5**, eaav1967 (2019).

[13] W.-C. Lo, C.-H. Fan, Y.-J. Ho, C.-W. Lin, and C.-K. Yeh, Tornado-inspired acoustic vortex tweezer for trapping and manipulating microbubbles, Proc. Natl. Acad. Sci. **118**, e2023188118 (2021).

[14] M. A. Ghanem, A. D. Maxwell, Y.-N. Wang, B. W. Cunitz, V. A. Khokhlova, O. A. Sapozhnikov, and M. R. Bailey, Noninvasive acoustic manipulation of objects in a living body, Proc. Natl. Acad. Sci. **117**, 16848 (2020).

[15] M. A. Ghanem, A. D. Maxwell, O. A. Sapozhnikov, V. A. Khokhlova, and M. R. Bailey, Quantification of acoustic radiation forces on solid objects in fluid, Phys. Rev. Appl. **12**, 044076 (2019).

[16] A. Marzo, M. Caleap, and B. W. Drinkwater, Acoustic virtual vortices with tunable orbital angular momentum for trapping of mie particles, Phys. Rev. Lett. **120**, 044301 (2018).

[17] K. Skeldon, C. Wilson, M. Edgar, and M. Padgett, An acoustic spanner and its associated rotational Doppler shift, New J. Phys. **10**, 013018 (2008).

[18] K. Volke-Sepulveda, A. O. Santillan, and R. R. Boullosa, Transfer of angular momentum to matter from acoustical vortices in free space, Phys. Rev. Lett. **100**, 024302 (2008).

[19] A. O. Santillán and K. Volke-Sepúlveda, A demonstration of rotating sound waves in free space and the transfer of their angular momentum to matter, Am. J. Phys. **77**, 209 (2009).

[20] C. E. Demore, Z. Yang, A. Volovick, S. Cochran, M. P. MacDonald, and G. C. Spalding, Mechanical evidence of the orbital angular momentum to energy ratio of vortex beams, Phys. Rev. Lett. **108**, 194301 (2012).

[21] L. Ye, C. Qiu, J. Lu, K. Tang, H. Jia, M. Ke, S. Peng, and Z. Liu, Making sound vortices by metasurfaces, AIP Adv. **6**, 085007 (2016).

[22] J. Liu, Z. Li, Y. Ding, A. Chen, B. Liang, J. Yang, J. C. Cheng, and J. Christensen, Twisting Linear to Orbital Angular Momentum in an Ultrasonic Motor, Adv. Mater. **34**, 2201575 (2022).

[23] Z. Hong, J. Zhang, and B. W. Drinkwater, Observation of orbital angular momentum transfer from Bessel-shaped acoustic vortices to diphasic liquid-microparticle mixtures, Phys. Rev. Lett. **114**, 214301 (2015).

[24] T. Wang, M. Ke, W. Li, Q. Yang, C. Qiu, and Z. Liu, Particle manipulation with acoustic vortex beam induced by a brass plate with spiral shape structure, Appl. Phys. Lett. **109**, 123506 (2016).

[25] Z. Hong, J. Yin, W. Zhai, N. Yan, W. Wang, J. Zhang, and B. W. Drinkwater, Dynamics of levitated objects in acoustic vortex fields, Sci. Rep. **7**, 7093 (2017).

[26] R. Zhang, H. Guo, W. Deng, X. Huang, F. Li, J. Lu, and Z. Liu, Acoustic tweezers and motor for living cells, Appl. Phys. Lett. **116**, 123503 (2020).

[27] L. Zhang and P. L. Marston, Acoustic radiation torque on small objects in viscous fluids and connection with viscous dissipation, J. Acoust. Soc. Am. **136**, 2917 (2014).

[28] A. D. Maxwell, M. Bailey, B. W. Cunitz, M. Terzi, A. Nikolaeva, S. Tsysar, and O. A. Sapozhnikov, Vortex beams and radiation torque for kidney stone management, J. Acoust. Soc. Am. **139**, 2040 (2016).

[29] X. Jiang, B. Liang, J. C. Cheng, and C. W. Qiu, Twisted Acoustics: Metasurface-Enabled Multiplexing and Demultiplexing, Adv. Mater. **30**, e1800257 (2018).

[30] C. Zhang, X. Jiang, J. He, Y. Li, and D. Ta, Spatiotemporal Acoustic Communication by a Single Sensor via Rotational Doppler Effect, Adv. Sci. **10**, e2206619 (2023).

[31] X. R. Li, D. J. Wu, Y. C. Luo, J. Yao, and X. J. Liu, Coupled Focused Acoustic Vortices Generated by Degenerated Artificial Plates for Acoustic Coded Communication, Adv. Mater. Technol. **7**, 2200102 (2022).

[32] H. Zhang and J. Yang, Transmission of video image in underwater acoustic communication, arXiv preprint arXiv:1902.10196 (2019).

[33] C. Shi, M. Dubois, Y. Wang, and X. Zhang, High-speed acoustic communication by multiplexing orbital angular momentum, Proc. Natl. Acad. Sci. **114**, 7250 (2017).

[34] J. L. Thomas and R. Marchiano, Pseudo angular momentum and topological charge conservation for nonlinear acoustical vortices, Phys. Rev. Lett. **91**, 244302 (2003).

[35] R. Marchiano and J. L. Thomas, Synthesis and analysis of linear and nonlinear acoustical vortices, Phys. Rev. E Stat. Nonlin. Soft Matter Phys. **71**, 066616 (2005).

[36] B. T. Hefner and B. R. Dzikowicz, A spiral wave front beacon for underwater navigation: basic concept and modeling, J. Acoust. Soc. Am. **129**, 3630 (2011).

[37] N. Jiménez, J. Ealo, R. D. Muelas-Hurtado, A. Duclos, and V. Romero-García, Subwavelength Acoustic Vortex Beams Using Self-Demodulation, Phys. Rev. Appl. **15**, 054027 (2021).

[38] S. Wang, G. Ma, and C. T. Chan, Topological transport of sound mediated by spin-redirection geometric phase, Sci. Adv. **4**, eaaq1475 (2018).

[39] F. Liu, W. Li, Z. Pu, and M. Ke, Acoustic waves splitter employing orbital angular Momentum, Appl. Phys. Lett. **114**, 193501 (2019).

[40] S. Gspan, A. Meyer, S. Bernet, and M. Ritsch-Marte, Optoacoustic generation of a helicoidal ultrasonic beam, J. Acoust. Soc. Am. **115**, 1142 (2004).

[41] J. L. Ealo, J. C. Prieto, and F. Seco, Airborne ultrasonic vortex generation using flexible ferroelectrets, IEEE Trans. Ultrason. Ferroelectr. Freq. Control **58**, 1651 (2011).

[42] X. Jiang, Y. Li, B. Liang, J. C. Cheng, and L. Zhang, Convert Acoustic Resonances to Orbital Angular Momentum, Phys. Rev. Lett. **117**, 034301 (2016).

[43] S.-W. Fan, Y.-F. Wang, L. Cao, Y. Zhu, A.-L. Chen, B. Vincent, B. Assouar, and Y.-S. Wang, Acoustic vortices with high-order orbital angular momentum by a continuously tunable metasurface, Appl. Phys. Lett. **116**, 163504 (2020).

[44] J. J. Liu, B. Liang, J. Yang, J. Yang, and J. C. Cheng, Generation of Non-aliased Two-dimensional Acoustic Vortex with Enclosed Metasurface, Sci. Rep. **10**, 3827 (2020).

[45] X. Jiang, J. Zhao, S.-l. Liu, B. Liang, X.-y. Zou, J. Yang, C.-W. Qiu, and J.-c. Cheng, Broadband and stable acoustic vortex emitter with multi-arm coiling slits, Appl. Phys. Lett. **108**, 203501 (2016).

[46] N. Jiménez, V. Sánchez-Morcillo, R. Pico, L. M. Garcia-Raffi, V. Romero-Garcia, and K. Staliunas, High-order acoustic Bessel beam generation by spiral gratings, Phys. Procedia **70**, 245 (2015).

[47] N. Jiménez, R. Picó, V. Sanchez-Morcillo, V. Romero-Garcia, L. M. Garcia-Raffi, and K. Staliunas, Formation of high-order acoustic Bessel beams by spiral diffraction gratings, Phys. Rev. E **94**, 053004 (2016).

[48] H. Zhou, J. Li, K. Guo, and Z. Guo, Generation of acoustic vortex beams with designed Fermat's spiral diffraction grating, J. Acoust. Soc. Am. **146**, 4237 (2019).

[49] D.-C. Chen, Q.-X. Zhou, X.-F. Zhu, Z. Xu, and D.-J. Wu, Focused acoustic vortex by an artificial structure with two sets of discrete Archimedean spiral slits, Appl. Phys. Lett. **115**, 083501 (2019).

[50] S. Jiménez-Gambín, N. Jiménez, J. M. Benlloch, and F. Camarena, Generating Bessel beams with broad depth-of-field by using phase-only acoustic holograms, Sci. Rep. **9**, 20104 (2019).

[51] S. Jiménez-Gambín, N. Jiménez, and F. Camarena, Transcranial focusing of ultrasonic vortices by acoustic holograms, Phys. Rev. Appl. **14**, 054070 (2020).

[52] A. S. Pilipchuk, A. A. Pilipchuk, and A. F. Sadreev, Generation of vortex waves in non-coaxial cylindrical waveguides, J. Acoust. Soc. Am. **146**, 4333 (2019).

[53] H. E. Hartig and C. E. Swanson, "Transverse" Acoustic Waves in Rigid Tubes, Phys. Rev. **54**, 618 (1938).

[54] M. Bruneau, *Fundamentals of acoustics* (John Wiley & Sons, 2013).

[55] F. Jacobsen and P. M. Juhl, *Fundamentals of general linear acoustics* (John Wiley & Sons, 2013).

[56] A. S. Pilipchuk, A. A. Pilipchuk, and A. F. Sadreev, Multi-channel bound states in the continuum in coaxial cylindrical waveguide, Phys. Scr. **94**, 115004 (2019).

[57] Z. Hou, H. Ding, N. Wang, X. Fang, and Y. Li, Acoustic vortices via nonlocal metagratings, Phys. Rev. Appl. **16**, 014002 (2021).

[58] J. Lekner, Acoustic beams with angular momentum, J. Acoust. Soc. Am. **120**, 3475 (2006).

[59] L. Zhang and P. L. Marston, Angular momentum flux of nonparaxial acoustic vortex beams and torques on axisymmetric objects, Phys. Rev. E **84**, 065601 (2011).